\documentclass[a4paper,11pt]{article}
\usepackage{pos}
\usepackage[english]{babel}
\usepackage[utf8]{inputenc}
\usepackage[T1]{fontenc}
\usepackage{pifont}
\usepackage{hyperref}
\usepackage{color,hyperref}
\usepackage{cleveref}
\usepackage{amsmath}
\usepackage{graphicx}
\usepackage{booktabs}
\usepackage{multirow}
\usepackage{subfigure}
\usepackage{float}
\usepackage{bm}

\newcommand{\beq}{\begin{eqnarray}}
\newcommand{\eeq}{\end{eqnarray}}
\newcommand{\be}{\begin{equation}}
\newcommand{\ee}{\end{equation}}
\newcommand{\ba}{\begin{eqnarray}}
\newcommand{\ea}{\end{eqnarray}}
\newcommand{\nn}{\nonumber}
\newcommand{\bit}{\begin{itemize}}
\newcommand{\eit}{\end{itemize}}
\newcommand{\rw}{\rightarrow}

\title{Lattice QCD calculation of the invisible decay $J/\psi \rw \gamma \nu\bar{\nu}$}
\author*[a]{Yu Meng}
\affiliation[a]{School of Physics and Microelectronics, Zhengzhou University, Zhengzhou, Henan 450001, China}

\emailAdd{yu\_meng@zzu.edu.cn}
\abstract{In this work, we present the first lattice QCD study on the invisible decay $J/\psi \rw \gamma\nu\bar{\nu}$. The calculation is accomplished using $N_f=2$ twisted mass fermion ensembles. The excited-state effects are observed and eliminated using a multi-state fit. The impact of finite-volume effects is also examined and confirmed to be well-controlled. After a continuous extrapolation under three lattice spacings, we obtain the branching fraction as $\operatorname{Br}[J/\psi \rw \gamma\nu\bar{\nu}]=1.00(9)(7)\times 10^{-10}$, where the first error is the statistical error and the second is an estimate of the systematics. The exact theoretical prediction can be used to remove the only invisible contamination from the standard model background in searching for the possible dark matter by the channel $J/\psi \rw \gamma+\textrm{invisible}$.}

\FullConference{%
 The 40th International Symposium on Lattice Field Theory, LATTICE2023
  31 July-4 August,2023 \\
Fermi National Accelerator Laboratory
}

\begin{document}
\maketitle

\section{Introduction}
Searching for dark matter is one of the major goals of contemporary astronomy and particle physics~\cite{Bertone:2004pz,Arkani-Hamed:2008hhe}. In recent decades, abundant experimental observations have hinted at the existence of dark matter, which triggered significant theoretical efforts to understand its nature and search for new physics beyond the Standard Model. Among various experimental detections, the heavy quarkonium experiments provide an ideal environment to study the possible dark matter associated with heavy quarks. In contrast to the low-energy dark matter nucleon scattering experiments, the decay of heavy quarkonium into a single photon and invisible particles can probe arbitrarily small dark matter masses. Therefore, it is widely used to search for light sterile neutrino or sub-GeV dark matter.

The CLEO~\cite{CLEO:2010juh}, BaBar~\cite{BaBar:2010eww}, Belle~\cite{Belle:2018pzt}, and BESIII~\cite{BESIII:2020sdo} experiments have performed the searches for $J/\psi$ or $\Upsilon$ radiative decays into invisible particles, and no signal was observed. The latest upper limits on the branching fraction of $J/\psi \rw \gamma+\textrm{invisible}$ is reported ranging from $8.3\times 10^{-8}$ to $1.8\times 10^{-6}$ by the BESIII experiment using $(2708.1\pm 14.5)\times 10^{6} \psi(3686)$ events collected by the detector~\cite{BESIII:2022rzz}. In this analysis, the invisible particle is interpreted as an axion-like particle(ALP), and the most stringent constraints on the ALP-photon coupling are presented. Not long before, the BESIII experiment also searches for a CP-odd light Higgs boson ($A^0$) in $J/\psi \rw \gamma A^0$~\cite{BESIII:2021ges}. Among these searches, the standard model decay $J/\psi \rw \gamma\nu\bar{\nu}$ is involved since the neutrinos are also invisible particles in the standard model. In Ref.~\cite{Gao:2014yga}, the author analyzes the process $J/\psi \rw \gamma\nu\bar{\nu}$ based on certain phenomenological assumptions and estimates the branching fraction as $\operatorname{Br}(J/\psi\rw \gamma\nu\bar{\nu})=0.7\times 10^{-10}$, thereby leaving a substantial room for new physics in the process. At present, several futural experiments are under planning or construction, such as Super Tau Charm Facility~\cite{Achasov:2023gey}, Belle II~\cite{Belle-II:2018jsg}, and LHCb~\cite{LHCb:2018roe}, have the great potential to significantly improve the upper limit on the branching fraction of $J/\psi \rw \gamma+\textrm{invisible}$.

At the present stage, a genuine non-perturbative calculation can not only provide a model-independent comparison with previous phenomenological studies but also provide a potential theoretical assist for experiments in the search for dark matter and new physics beyond the standard model. In this paper, we present the first lattice calculation of the invisible decay $J/\psi \rw \gamma+\textrm{invisible}$. The aim of the work is to non-perturbatively determine the branching fraction with various systematic effects under well control.
\begin{figure}[!h]
\centering
\subfigure{\includegraphics[width=0.5\textwidth]{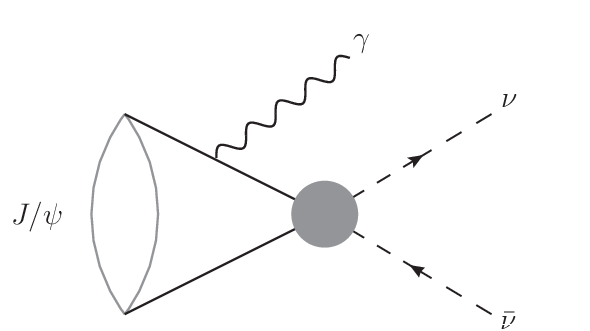}}\hspace{5mm}
\caption{The diagram for the decay $J/\psi\rw \gamma\nu\bar{\nu}$, where the shaded region denotes a weak neutral current.}
\label{fig:diagram}
\end{figure}

\subsection{Foundation}\label{sec:foundation}

\section{Approach to the decay width on the lattice}\label{sec:method}

We start our discussion from the amplitude of $J/\psi\rw \gamma\nu\bar{\nu}$, the lowest-order contribution of which is expressed by
\beq\label{eq:amp}
\mathcal{M}=H_{\mu\nu\alpha}(q,p)\epsilon^{\alpha}_{J/\psi}(p)(-ie\epsilon^{\nu*}(q))(-\frac{i}{2}g_Z)^2 
\times
\bar{u}(q_1)\frac{\gamma^{\mu}}{2}(1-\gamma_5)v(q_2)\frac{-i}{(k^2-m_Z^2)}
\eeq
where the nonperturbative hadronic interaction between the $J/\psi$, photon and $Z$ boson is encoded in a hadronic function $H_{\mu\nu\alpha}(q,p)$,
\beq\label{eq:hadron_mom}
H_{\mu\nu\alpha}(q,p)=\int d^4x\textrm{e}^{iqx}\mathcal{H}_{\mu\nu\alpha}(x,p)
\eeq
where the hadronic function $\mathcal{H}_{\mu\nu\alpha}(x,p)$ is defined as
\be\label{eq:hadron_function_x}
\mathcal{H}_{\mu\nu\alpha}(x,p)=\langle 0|T\{J_{\mu}^{\textrm{em}}(x)J_{\nu}^{Z}(0)\}|J/\psi(p)_{\alpha} \rangle
\ee
with $J/\psi$ four-momentum $p=(m_{J/\psi},\vec{0})$, photon $q=(|\vec{q}|,\vec{q})$ and the neutrino $q_i=(|\vec{q}_i|,\vec{q}_i)$,$i=1,2$.
Both the photons and neutrinos satisfy the on-shell conditions and are viewed as massless. The electromagnetic and weak currents are defined as $J_\mu^{\textrm{em}}=\sum_qe_q\,\bar{q}\gamma_\mu q$($e_q=2/3,-1/3,-1/3,2/3$ for $q=u,d,s,c$), $J_\nu^{Z}=\sum_q \bar{q}\gamma_\nu(g_V^q-g_A^q\gamma_5) q$, $g_V^{q}=T_3^{q}-2e_q\sin^2\theta_W$ and $g_A^q=T_3^q$, where $T_3^q$ is the third component of the weak isospin of the fermion. In the case of the charm quark, we know $g_A^c=1/2$ and $g_V^c=1/2-4/3\sin^2\theta_W$.
The $\epsilon_{J/\psi}^{\alpha}(p)$ is the polarization vector of $J/\psi$ and $\epsilon^{\nu}(q)$ for the photon. The $e$ is the coupling constant of electromagnetic interaction, and $g_Z$ depicts the coupling of $Z$ boson to the fermions. The $Z$ boson mass is $m_Z$ and the four-momentum is given by $k=q_1+q_2$.

For the virtual $Z$ boson, $k^2 \ll m_Z^2 $, it is natural to make an replacement for the $Z$ boson propagator
\be
\frac{1}{k^2-m_Z^2}\rw -\frac{1}{m_Z^2}
\ee

Also considering the following notations,
\be
\frac{G_F}{\sqrt{2}}=\frac{g_W^2}{8m_W^2}, g_Z=\frac{g_W}{\cos\theta_W}, \cos\theta_W=\frac{m_W}{m_Z}
\ee
The amplitude in Eq.~(\ref{eq:amp}) thereby reduces to
\beq\label{eq:amp2}
\mathcal{M}=-e\frac{G_F}{\sqrt{2}}H_{\mu\nu\alpha}(q,p)\epsilon^{\alpha}_{J/\psi}(p)\epsilon^{\nu*}(q)
\times\bar{u}(q_1)\gamma^{\mu}(1-\gamma_5)v(q_2)
\eeq
With consideration of the gauge symmetry and parity, the hadronic function $H_{\mu\nu\alpha}(q,p)$ can be parameterized as~\cite{Gao:2014yga}
\be\label{eq:form_factor}
H_{\mu\nu\alpha}(q,p)\equiv \epsilon_{\mu\nu\alpha\beta}q_{\beta}F_{\gamma\nu\bar{\nu}}
\ee
The direct calculation on the decay width of $J/\psi \rw \gamma\nu\bar{\nu}$ in the rest frame of $J/\psi$, by employing Eq.~(\ref{eq:amp2}) and (\ref{eq:form_factor}), leads to
\beq\label{eq:width}
\Gamma(J/\psi\rw \gamma\nu\bar{\nu}) &=&\frac{1}{2m_{J/\psi}}\int\frac{d^3\vec{q}}{(2\pi)^32|\vec{q}|}\int\frac{d^3\vec{q_1}}{(2\pi)^32|\vec{q}_1|}\int\frac{d^3\vec{q}_2}{(2\pi)^32|\vec{q}_2|} \nn \\
&\times&(2\pi)^4\delta^4(p-q-q_1-q_2)\times \frac{1}{3}|\mathcal{M}|^2 \times 3 \nn \\
&=&\frac{\alpha G_{F}^2}{3\pi^2}\int_{0}^{\frac{m_{J/\psi}}{2}} |\vec{q}|^3(m_{J/\psi}-|\vec{q}|)|F_{\gamma\nu\bar{\nu}}|^2 d|\vec{q}|
\eeq
where $\alpha\equiv e^2/4\pi$. Factor 1/3 in the third line denotes the average of three polarizations of $J/\psi$ in its rest frame and factor 3 for the three flavors of neutrinos.

\subsection{Relationship of hadronic function in Minkowski and Euclidean space}\label{sec:relation_M_E}
In this section, we present the relation between the hadronic functions in Minkowski and Euclidean spacetime, which can be established by inserting a complete set of intermediate states into the respective hadronic functions.

In the Minkowski spacetime, the hadronic function has the following decomposition
\beq\label{eq:H_expand_M}
H_{\mu\nu\alpha}(q,p)
&=&i\sum\limits_{n,\vec{q}}\frac{1}{E_{\gamma}-E_{n}+i\epsilon}\langle 0|J_{\mu}^{\textrm{em}}(0)|n(\vec{q})\rangle \langle n(\vec{q})|J_{\nu}^{Z}(0)|J/\psi(p)_{\alpha} \rangle \nonumber \\
&-&i\sum\limits_{n',\vec{q}}\frac{1}{E_{\gamma}+E_{n'}-m_{J/\psi}-i\epsilon} \langle 0|J_{\nu}^{Z}(0)|n'(-\vec{q})\rangle \langle n'(-\vec{q})|J_{\mu}^{\textrm{em}}(0)|J/\psi(p)_{\alpha} \rangle \nonumber \\
\eeq
where the first line corresponds to the time-ordering $t>0$ and second line for $t<0$ in Eq.~(\ref{eq:hadron_mom}). The intermediate states $|n\rangle$ and $|n'\rangle$ represent all possible states with the allowed quantum numbers. As far as the connected contribution is concerned in this work, the low-lying states are given by $|n'\rangle=|J/\psi\rangle$ and $|n\rangle=|\eta_c\rangle$, respectively.

In the Euclidean spacetime, the hadronic function in Eq.~(\ref{eq:hadron_mom}) is replaced by $H_{\mu\nu\alpha}^E(q,p)$, which is obtained by making a naive Wick rotation $t\rw -it$
\beq
H_{\mu\nu\alpha}^E(q,p)&=&-i\int_{-T/2}^{T/2} dt \int d^3\vec{x} \textrm{e}^{E_{\gamma}t-i\vec{q}\cdot \vec{x}}\mathcal{H}_{\mu\nu\alpha}(x,p) \nonumber \\
\eeq
with the Euclidean momenta $q=(iE_{\gamma},\vec{q}),p=(im_{J/\psi},0)$.
As before, after inserting a complete set of intermediate states into the Euclidean hadronic function above, we obtain
\beq\label{eq:H_expand_E}
H_{\mu\nu\alpha}^E(q,p)
&=&i\sum\limits_{n,\vec{q}}\frac{1-\textrm{e}^{-(E_n-E_{\gamma})T/2}}{E_{\gamma}-E_{n}+i\epsilon}\langle 0|J_{\mu}^{\textrm{em}}(0)|n(\vec{q})\rangle \langle n(\vec{q})|J_{\nu}^{Z}(0)|J/\psi(p)_{\alpha} \rangle \nonumber \\
&-&i\sum\limits_{n',\vec{q}}\frac{1-\textrm{e}^{-(E_{\gamma}+E_{n'}-m_{J/\psi})T/2}}{E_{\gamma}+E_{n'}-m_{J/\psi}-i\epsilon} \langle 0|J_{\nu}^{Z}(0)|n'(-\vec{q})\rangle \langle n'(-\vec{q})|J_{\mu}^{\textrm{em}}(0)|J/\psi(p)_{\alpha} \rangle \nonumber \\
\eeq
where the finite time integral $[-T/2,T/2]$ is introduced to define the Euclidean hadronic function.

Whether the Minkowski hadronic function can be obtained from the Euclidean hadronic function by naive Wick rotation usually depends on whether all the $T$-dependence terms converge in the limit $T\rw \infty$. If it does, the Wick rotation will leave the hadronic function unchanged and the lattice calculation produces the physical results without particular difficulties. In this study, it requires the conditions
\be\label{eq:condition_1}
E_n-E_{\gamma}>0
\ee
\be\label{eq:condition_2}
E_{\gamma}+E_{n'}-m_{J/\psi}>0
\ee
must be satisfied for $E_{\gamma} \in [0,m_{J/\psi}/2]$.

For the time ordering $t>0$, where the weak current is inserted before the electromagnetic current, the low-lying state is $J/\psi$ particle with momentum $\vec{q}$ and the condition (\ref{eq:condition_1}) is satisfied readily. However, the situation is quite different for time ordering $t<0$, where the electromagnetic current is inserted before the weak current. In this case, the low-lying state is $\eta_c$ particle whose mass is slightly less than the initial state $J/\psi$, resulting in a violation of condition (\ref{eq:condition_2}) for very small $E_{\gamma}$, for example, $E_{\gamma}=0$. For all ensembles used in this work, we find there exists only one momentum $\vec{q}=0$ for intermediate state $|\eta_c(\vec{q})\rangle$ that violates the condition (\ref{eq:condition_2}), leading to an exponentially growing factor $e^{-(E_{\gamma}+E_{n'}-m_{J/\psi})T/2}$ as $T$ increases. One can check it numerically using the discrete energy levels of $\eta_c$ summarized in Table~\ref{tab:disper}. Moreover, for $\vec{q}=0$ it has
\be
\langle 0|J_{\nu}^{Z}(0)|\eta_c(\vec{0})\rangle \langle \eta_c(\vec{0})|J_{\mu}^{\textrm{em}}(0)|J/\psi(\vec{0})_{\alpha} \rangle=0
\ee
which still protects the Euclidean hadronic function from an exponentially growing factor $e^{-(m_{\eta_c}-m_{J/\psi})T/2}$. In other words, all the intermediate states with discrete momenta $\vec{q}=2\pi\vec{n}/L$ are independent of the $T$-dependence factor as $T\rw \infty$. We conclude that for the time ordering $t<0$, the condition (\ref{eq:condition_1}) is also satisfied in our calculations. Thus, we have proved that one can extract the Minkowski hadronic function from the Euclidean hadronic function directly with naive Wick rotation, and the $i\epsilon$ in Eq.(\ref{eq:H_expand_M}) and Eq.(\ref{eq:H_expand_E}) are unnecessary.

\subsection{Extraction of the hadroinc function from lattice data}\label{sec:had_lat}
In the above section, we have established the direct connection between the
Minkowski hadronic function and the Euclidean hadronic function. In the following, we will provide the details of constructing the Euclidean hadronic function using the lattice data.

The hadronic function $\mathcal{H}_{\mu\nu\alpha}(x,p)$ defined in Eq.(\ref{eq:hadron_function_x}) can be extracted from a three-point function $C_{\mu\nu\alpha}^{(3)}(x;\Delta t)$
\be
\label{eq:decay_width}
C_{\mu\nu\alpha}^{(3)}(x;\Delta t)=
\left\{
\begin{array}{lr}
\langle J_{\mu}^{\textrm{em}}(x)J_{\nu}^{Z}(0)\phi^{\dagger}_{J/\psi,\alpha}(-\Delta t) \rangle, &t\geq 0 \\
\langle J_{\mu}^{Z}(0) J_{\nu}^{\textrm{em}}(x)\phi^{\dagger}_{J/\psi,\alpha}(t-\Delta t) \rangle, &t<0 \\
\end{array}
\right.
\ee
where $\phi_{J/\psi,\alpha}$ is the $J/\psi$ interpolating operator. A sufficient large $\Delta t$ should be chosen to guarantee $J/\psi$ ground-state dominance. For a finite $\Delta t$, the hadronic function has a $\Delta t$ dependence, we thereby denote the hadronic function $\mathcal{H}_{\mu\nu\alpha}(x,p)$ as $\mathcal{H}_{\mu\nu\alpha}(x,\Delta t)$ where the initial momentum $p$ is omitted since our calculation is limited to the rest frame. So, it has
\be
\mathcal{H}_{\mu\nu\alpha}(x,\Delta t)=
\left\{
\begin{array}{lr}
\frac{2m_{J/\psi}}{Z_0}\textrm{e}^{m_{J/\psi}\Delta t} C^{(3)}_{\mu\nu\alpha}(x;\Delta t), &t\geq 0 \\
\frac{2m_{J/\psi}}{Z_0}\textrm{e}^{m_{J/\psi}(\Delta t-t)} C^{(3)}_{\mu\nu\alpha}(x;\Delta t), &t<0 \\
\end{array}
\right.
\ee
with $Z_0=\langle J/\psi | \phi_{J/\psi}^{\dagger}|0\rangle$ the overlap amplitude for the $J/\psi$ ground state. Both $Z_0$ and $m_{J/\psi}$ can be calculated from the two-point function $C^{(2)}(t)=\langle \phi_{J/\psi}(t)
\phi^{\dagger}_{J/\psi}(0)\rangle$, which has the following expression
\be\label{eq:2pt}
C^{(2)}(t)=\sum_{i=0,1}\frac{Z_i^2}{2E_i} \left(\textrm{e}^{-E_it}+\textrm{e}^{-E_i(T-t)}\right)
\ee
We adopt a two-state fit form for the two-point function $C^{(2)}(t)$ to extract $Z_i,E_i(i=1,2)$, with $E_0=m_{J/\psi}$ the ground state energy, $E_1$ the energy of the first excited state and $Z_1$ the overlap amplitude for the first excited state.
As is pointed out in the previous paper, when the precision reaches a few percent in our calculation, the excited-state effects are statistically significant unless $t \gtrsim 1.6$ fm as far as $C^{(2)}(t)$ is concerned. Such systematic effects also affect the three-point function $C^{(3)}_{\mu\nu\alpha}$, leading to an obvious $\Delta t$ dependence. In a realistic lattice calculation, a series of $\Delta t$ are utilized to perform an infinite extrapolation $\Delta t\rw \infty$.

\subsection{Form factor and decay width}\label{sec:F_width}
To compute the $F_{\gamma\nu\bar{\nu}}$, the traditional way is to choose a series of lattice momenta $\vec{q}=2\pi\vec{n}/L$ with $\vec{n}=[0 0 1],[0 1 1],[1 1 1] \cdots$, and the phase space integral is finally completed by interpolating or fitting this discrete $ F_{\gamma\nu\bar{\nu}}(|\vec{q}|)$, leading to a model-dependent systematic effect. In this work, we will proceed to another way, which is widely called the scalar function method. The method has been widely
applied to various processes~\cite{Feng:2020zdc,Ma:2021azh,Tuo:2021ewr,Meng:2021las,Fu:2022fgh}. The key point of the method is to construct the appropriate scalar function method to extract the relevant form factors with particular momenta. The decay width, which is related to the form factors directly by the phase-space integral, can be calculated using the Monte-Carlo method.

According to the parameterization of the hadronic function $H_{\mu\nu\alpha}(q,p)$ in Eq.~(\ref{eq:form_factor}), we construct the scalar function $\mathcal{I}$ by multiplying $\epsilon_{\mu\nu\alpha\beta}p_{\beta}$ to both sides. After averaging over the direction of $\vec{q}$, it arrives at
\beq\label{eq:I_function}
\mathcal{I}(E_{\gamma},\Delta t) 
=im_{J/\psi}\int e^{E_{\gamma}t}dt\int d^3\vec{x}j_0(E_{\gamma}|\vec{x}|)\epsilon_{\mu\nu\alpha 0}\mathcal{H}_{\mu\nu\alpha}(x,\Delta t) \nonumber \\
\eeq
where $E_{\gamma}\equiv |\vec{q}|$. Then the form factor is extracted through
\be
F_{\gamma\nu\bar{\nu}}(E_{\gamma},\Delta t)= -\frac{1}{6m_{J/\psi}E_{\gamma}}\mathcal{I}(E_{\gamma},\Delta t)
\ee

Using the form factor $F_{\gamma\nu\bar{\nu}}(E_{\gamma},\Delta t)$ as input, the decay width of $J/\psi \rw \gamma\nu\bar{\nu}$ can be obtained by the Monte-Carlo phase-integral in the region $E_{\gamma}\in [0,m_{J/\psi}/2]$
\beq\label{eq:decay_width_MC}
\Gamma_{\gamma\nu\bar{\nu}}(\Delta t) 
=\frac{\alpha G_{F}^2}{3\pi^2}\frac{m_{J/\psi}}{2N_{MC}}\sum\limits_{i=1}^{N_{MC}}\left(E_{\gamma}^3(m_{J/\psi}-E_{\gamma})|F_{\gamma\nu\bar{\nu}}(E_{\gamma},\Delta t)|^2 \right)_i \nonumber \\
\eeq
where $N_{MC}$ is the number of Monte-Carlo simulations, which is chosen to guarantee the Monte-Carlo error is much less than the statistical error.

To further reduce the lattice discretization effect, we define a dimensionless quantity $R_f\equiv \Gamma_{\gamma\nu\bar{\nu}}/f_{J/\psi}$, where $f_{J/\psi}$ is the decay constant of $J/\psi$. The $\Delta t$ dependence can be parameterized using a relatively simple two-state form
\be\label{eq:th_fit}
R_f(\Delta t)=R_f+\zeta \cdot \textrm{e}^{-(E_1-E_0)\Delta t}
\ee
with two unknown parameters $R_f$ and $\zeta$. After the continuous extrapolations for the dimensionless $R_f$ and the decay constant $f_{J/\psi}$, we obtain the physical results as $R_f^{\textrm{Cont.Limit}}$ and $f_{J/\psi}^{\textrm{Cont.Limit}}$. The physical decay width can be therefore obtained by rescaling $R_f^{\textrm{Cont.Limit}}$ after multipling $f_{J/\psi}^{\textrm{Cont.Limit}}$. Finally, the branching fraction is presented as follows
\be
\operatorname{Br}[J/\psi \rw \gamma\nu\bar{\nu}]=R_f^{\textrm{Cont.Limit}}\times \frac{f_{J/\psi}^{\textrm{Cont.Limit}}}{\Gamma_{J/\psi}}
\ee
where $\Gamma_{J/\psi}=92.6$ keV is the $J/\psi$ decay width from the Particle Data Group.

\section{Numerical setup}\label{sec:setup}

\begin{table}[!h]
\centering
\begin{tabular}{cccccc}
\hline 
\hline
\textrm{Ensemble} & $a$ (fm) & $L^3\times T$ & $N_{\textrm{conf}}\times T$
& $m_{\pi} (\textrm{MeV})$ & $t$ \\
\hline
a67 & 0.0667(20) & $32^3\times 64$& $197\times 64$ & 300 & 12-18 \\
a85 & 0.085(2) & $24^3\times 48$ & $200\times 48$ & 315 & 10-14 \\
a98 & 0.098(3) & $24^3\times 48$ & $236\times 48$ & 365 & 9-13 \\
\hline
\end{tabular}
\caption{
Parameters of gauge ensembles are used in this work. From left to right, we list the ensemble name, the lattice spacing $a$,
the spatial and temporal lattice size $L$ and $T$, the number of the measurements
of the correlation function for each ensemble $N_{\textrm{conf}}\times T$ with $N_{\textrm{conf}}$ the number of the configurations
used , the pion mass $m_{\pi}$ and the range of the time separation $t$ between the initial hadron and the electromagnetic current.
Here, both $L$, $T$ and $t$ are given in lattice units.}\label{table:cfgs}
\end{table}

We use three two-flavor twisted mass gauge ensembles generated by
the Extended Twisted Mass Collaboration (ETMC)~\cite{ETM:2009ptp,Becirevic:2012dc} with lattice spacing
$a \simeq 0.0667,0.085,0.098$ fm, respectively. For convenience, we name them a67, a85, and a98 in this work.
The ensemble parameters are shown in Table.~\ref{table:cfgs}. The valence charm quark mass is tuned
by setting the lattice result of $J/\psi$ mass to the physical one.
The detailed information on the tuning is referred to Ref.~\cite{Meng:2021ecs}.

In this work, we calculate the three-point correlation function
$C^{(3)}_{\mu\nu\alpha}(\vec{x},t)$ using $Z_4$-stochastic wall-source $J/\psi$ interpolating operator $\phi_{J/\psi,\alpha}=\bar{c}\gamma_{\alpha}c$. For time ordering $t\geq 0$, we place the point source propagator on $J_{\nu}^{Z}$ and treat the electromagnetic current $J_{\mu}^{\textrm{em}}$ as the sink. For the time ordering $t<0$, after considering the space-time translation invariance of the correlation function, i.e.
$ \langle J_{\mu}^{Z}(0) J_{\nu}^{\textrm{em}}(x)\phi^{\dagger}_{J/\psi,\alpha}(t-\Delta t) \rangle =\langle J_{\mu}^{Z}(-\vec{x},-t) J_{\nu}^{\textrm{em}}(0)\phi^{\dagger}_{J/\psi,\alpha}(-\Delta t) \rangle$, we place the point source propagator on $J_{\nu}^{\textrm{em}}$ and treat the weak current $J_{\mu}^{Z}$ as the sink. The wall-source propagator used here can able to reduce the uncertainty of the mass spectrum by nearly half. All the propagators are produced on all time slices by average to increase the statistics based on time translation invariance. We also apply the APE~\cite{APE:1987ehd} and Gaussian smearing~\cite{Gusken:1989qx} to the $J/\psi$ field to efficiently reduce the excited-state effects.

To compute the $f_{J/\psi}$, we calculate the two-point function $C_{ii}^{(2)}(t)=\langle \mathcal{O}_{i}(t)\mathcal{O}_{i}^{\dagger}(0)\rangle$ using a point source $J/\psi$ interpolating operator $\mathcal{O}_{i}=Z_A\bar{c}\gamma_{i}c$. The overlap amplitude $Z_{0i}=\langle 0|\bar{c}\gamma_{i}c(0)|J/\psi(\vec{0},\lambda)\rangle$ can be extracted from a simple single-state fit
\be
C_{ii}^{(2)}(t)=\frac{Z_A^2Z_{0i}^2}{2m_{J/\psi}}\left(\textrm{e}^{-m_{J/\psi}t}+\textrm{e}^{-m_{J/\psi}(T-t)}\right)
\ee
then the $J/\psi$ decay constant is obtained immediately by $f_{J/\psi}=Z_AZ_{0i}/m_{J/\psi}$.

In our calculations, we choose the local vector current $J_\nu^{\textrm{em}}(x)=Z_V e_c \bar{c}\gamma_{\nu}c$ and weak current $J_{\mu}^{Z}=\bar{c}\gamma_{\mu}(Z_Vg_V^c-Z_Ag_A^c\gamma_5)c$, where the renormalization constants $Z_V$ and $Z_A$ are introduced. The detailed determination of $Z_V$ has been presented in our previous paper~\cite{Meng:2021ecs}.
In this study, we just quote the values directly, which are shown as 0.6047(19), 0.6257(21), and 0.6516(15) for $a=0.098,0.085,0.0667$ fm, respectively. The values of $Z_A$ are referred to the paper~\cite{ETM:2010iwh}, which are calculated by the RI-MOM scheme, and the results are given as 0.746(11),0.746(06) and 0.772(06) for $a=0.098,0.085,0.0667$ fm, respectively.

\section{Numerical results}\label{sec:result}

\subsection{Check of condition (\ref{eq:condition_2})}\label{sec:check}
\begin{table}[!h]
\center
\begin{tabular}{cccccc}
\hline
\hline
Ensemble & a67 & a85 &a98 \\
\hline
$aE_{\eta_c}(|\vec{n}|^2=0)$ & 1.0142(2) & 1.2958(3) & 1.4995(3) \\
$aE_{\eta_c}(|\vec{n}|^2=1)$ & 1.0302(2) & 1.3157(3) & 1.5144(4) \\
$aE_{\eta_c}(|\vec{n}|^2=2)$ & 1.0467(2) & 1.3354(3) & 1.5290(4) \\
$aE_{\eta_c}(|\vec{n}|^2=3)$ & 1.0629(3) & 1.3546(4) & 1.5434(4) \\
$aE_{\eta_c}(|\vec{n}|^2=4)$ & 1.0782(4) & 1.3729(5) & 1.5572(5) \\
\hline
$a\delta E(|\vec{n}|^2=0)$ & -0.0343(2) & -0.0372(3) & -0.0387(3) \\
$a\delta E(|\vec{n}|^2=1)$ & 0.1781(2) & 0.2446(3) & 0.2380(4) \\
$a\delta E(|\vec{n}|^2=2)$ & 0.2758(3) & 0.3737(3) & 0.3611(4) \\
$a\delta E(|\vec{n}|^2=3)$ & 0.3544(3) & 0.4751(4) & 0.4587(4) \\
$a\delta E(|\vec{n}|^2=4)$ & 0.4223(4) & 0.5636(5) & 0.5426(5) \\
\hline
\end{tabular}
\caption{Numerical results of $E_{\eta_c}(\vec{p})$ and $\delta E(\vec{p})$ with $\vec{p}=2\pi\vec{n}/L,|\vec{n}|^2=0,1,2,3,4$.}
\label{tab:disper}
\end{table}
In Sec~\ref{sec:relation_M_E}, we have declared that $\delta E(\vec{p})=|\vec{p}|+E_{\eta_c}(\vec{p})-m_{J/\psi}>0$ is valid for any non-zero lattice momentum $\vec{p}=2\pi\vec{n}/L$, so the condition (\ref{eq:condition_2}) is satisfied in our work. Using a point-source propagator, we extract a series of discrete energy levels of $\eta_c$ from the two-point function calculated by the interpolating operator $\mathcal{O}_{\eta_c}=\bar{c}\gamma_5 c$. The numerical values of $E_{\eta_c}(\vec{p})$ and $\delta E(\vec{p})$ are also summarized in Table~\ref{tab:disper}. It is shown readily that $\delta E(\vec{p})>0$ for $|\vec{n}|^2\neq 0$, hence leading to a guarantee of condition (\ref{eq:condition_2}).

\subsection{$f_{J/\psi}$}\label{sec:f_decay}

\begin{figure}[!h]
\centering
\subfigure{\includegraphics[width=0.6\textwidth]{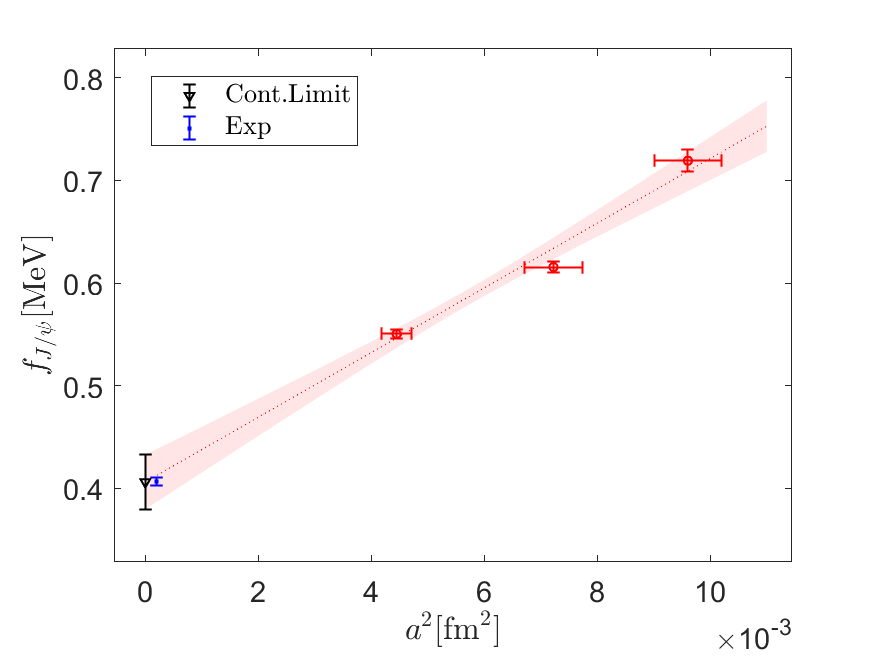}}\hspace{5mm}
\caption{Lattice results of $f_{J/\psi}$ as a function of lattice spacing. The errors of lattice spacing are included
in the fitting and presented by the horizontal error bars. The symbol of the red circle denotes the lattice results
from ensemble a67,a85, and a98 from left to right. The black triangle is the result in continuous limit $a^2\rw 0$ and circle blue is obtained using the experimental average of $\Gamma_{e+e-}$ and $\alpha_{QED}(m_{J/\psi}^2)=1/134.02$.}
\label{fig:f_jpsi_cont_limit}
\end{figure}

We present the lattice results of the decay constant $f_{J/\psi}$ in different lattice spacings in Fig.~\ref{fig:f_jpsi_cont_limit}. The continuous extrapolation which is linear in $a^2$ is performed due to the so-called automatic $\mathcal{O}(a)$ improvement for the twisted mass configuration. After the continuous extrapolation, we obtain
\be
f_{J/\psi}^{\textrm{Cont.Limit}}=406(26) \,\textrm{MeV}
\ee

Our lattice result is consistent with the experimental result $f_{J/\psi}^{\textrm{exp}}=406.5(3.7)$ MeV but with a larger statistical error.
The experimental value is obtained using the experimental average of $\Gamma_{e^+e^-}$ and $\alpha_{QED}(m_{J/\psi})$ through
\be
\Gamma_{e^+e^-}=\frac{4\pi}{3}\alpha^2_{QED}(M_{J/\psi}^2)e_c^2\frac{f_{J/\psi}^2}{M_{J/\psi}}
\ee
where $\alpha_{QED}(M_{J/\psi}^2)$ is evaluated at the scale of $M_{J/\psi}=3096.9$ MeV. Note that the latest lattice QCD calculation from HPQCD~\cite{Hatton:2020qhk} gives a value $f_{J/\psi,QCD}=409.6(1.6)$ with a much smaller statistical error than this work.

\subsection{Finite-volume effects} \label{sec:FV}
The decay width is calculated by a Monte-Carlo phase-integral as showed in Eq.~(\ref{eq:decay_width_MC}), where $N_{MC}=200$ is chosen and examined to guarantee the phase-integral error is much less than the statistical error. In our calculations, the integral energy $E_{\gamma} \in [0,m_{J/\psi}/2]$ is picked randomly.
The non-lattice values ($E_{\gamma}\neq 2\pi|\vec{n}|/L$) will inevitably introduce the systematic effects. These effects are essentially the finite-volume effects, since all the random values $E_{\gamma}$ become the lattice ones as the volume $L$ goes to infinity.

\begin{figure}[!h]
\centering
\subfigure{\includegraphics[width=0.6\textwidth]{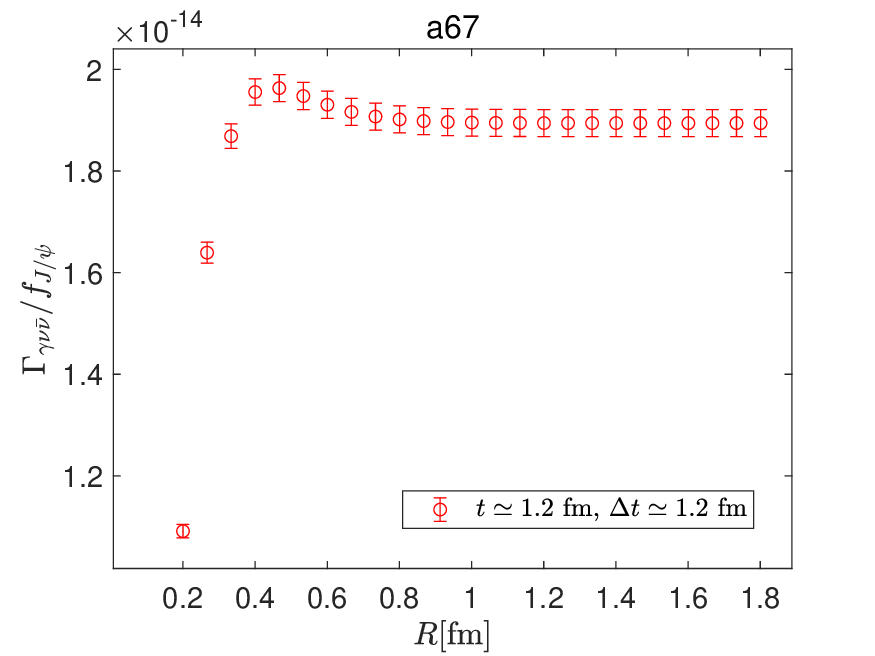}}\hspace{5mm}
\caption{For ensemble a67, $\Gamma_{\gamma\nu\bar{\nu}}/f_{J/\psi}$ with $t\simeq 1.2$ fm and $\Delta t \simeq 1.2$ fm as a function of the spatial range truncation $R$.}
\label{fig:R-dependence}
\end{figure}

To examine the finite-volume effects, we introduce a spatial integral truncation parameter $R$ in Eq.~(\ref{eq:I_function}).
As the hadronic function $\mathcal{H}_{\mu\nu\alpha}(x)$ is dominated by the $\eta_c$ state at large $|\vec{x}|$, the size of the integrand is exponentially suppressed when $|\vec{x}|$ becomes large.
In Fig.~\ref{fig:R-dependence} the ratio $\Gamma_{\gamma\nu\bar{\nu}}/f_{J/\psi}$ is shown as
a function of $R$. It is clearly seen that there exists a plateau for $R\gtrsim 0.8$ fm,
indicating that the hadronic function $\mathcal{H}_{\mu\nu\alpha}(x)$
at $|\vec{x}|\gtrsim 0.8$ fm has negligible contribution to $\Gamma_{\gamma\nu\bar{\nu}}/f_{J/\psi}$.
All the ensembles have the lattice size $L>2$ fm which is sufficiently large to accommodate the hadron.
We thus conclude that finite-volume effects are well under control in our calculation.

\subsection{Decay width}\label{sec:final_result}

\begin{figure}[!h]
\centering
\subfigure{\includegraphics[width=0.6\textwidth]{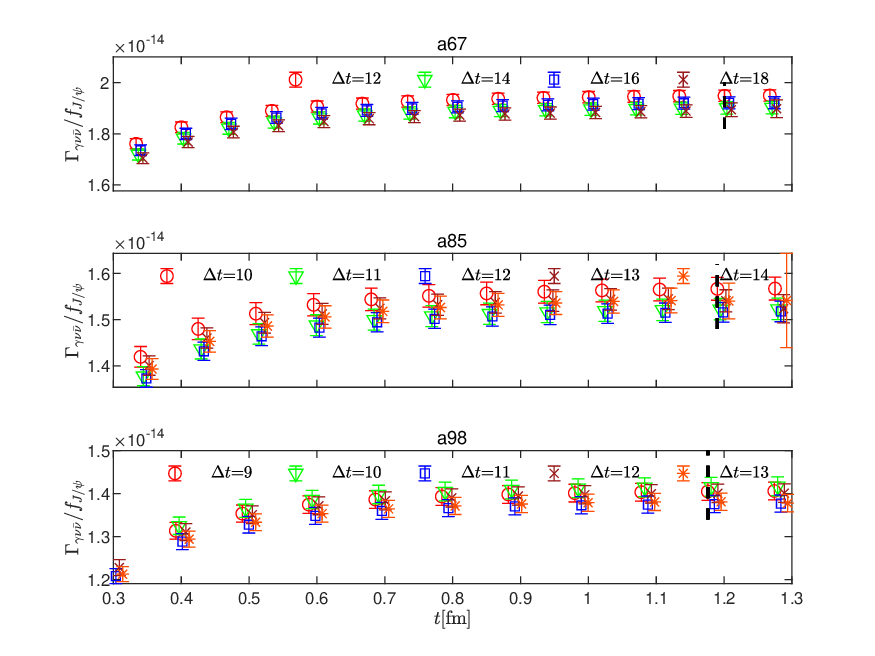}}\hspace{5mm}
\caption{The lattice results of $\Gamma_{\gamma\nu\bar{\nu}}/f_{J/\psi}$ for ensemble a67, a85 and a98, which are shown as a function of $t$ with various choices of $\Delta t$. The vertical dashed line denotes a conservative choice of $t \simeq 1.2$ fm, where the ground-state saturation is realized. The statistical error of $Z_A$ is not included here.}
\label{fig:F_th}
\end{figure}

\begin{figure}[!h]
\centering
\subfigure{\includegraphics[width=0.6\textwidth]{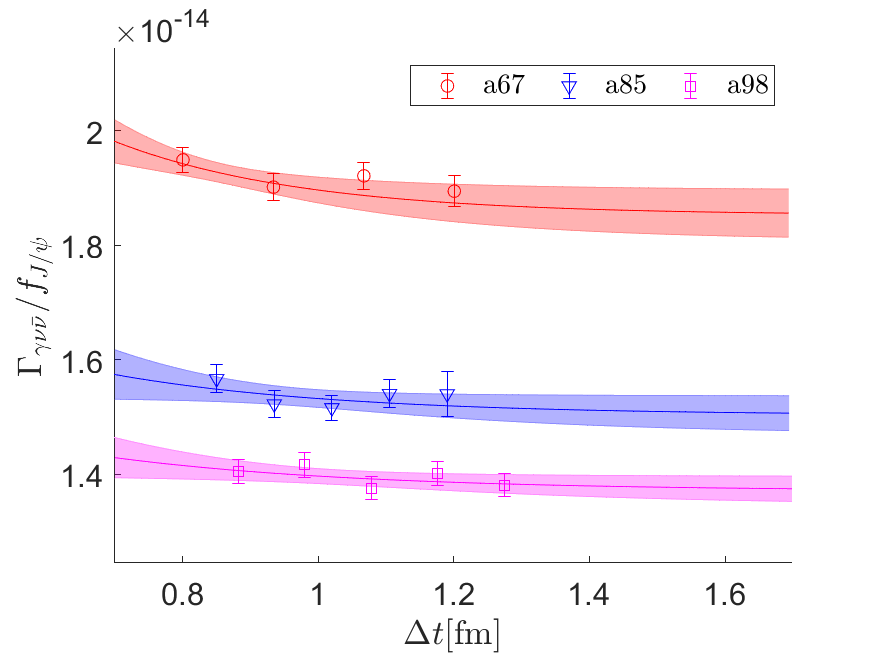}}\hspace{5mm}
\caption{The lattice results of $\Gamma_{\gamma\nu\bar{\nu}}/f_{J/\psi}$ with the cut $t\simeq 1.2$ fm in Fig.\ref{fig:F_th} are shown as a function of $\Delta t$ together with a fit to the form~(\ref{eq:th_fit}).}
\label{fig:F_th_limit}
\end{figure}

The lattice results of $\Gamma_{\gamma\nu\bar{\nu}}/f_{J/\psi}$ as a function of $t$ with different seperation $\Delta t$ are shown in Fig.~\ref{fig:F_th}. We find that for all the separation $\Delta t$ and all ensembles used in this work,
a temporal truncation $t\simeq 1.2$ fm is a conservative choice for the ground-state saturation. With this choice, the results for $\Gamma_{\gamma\nu\bar{\nu}}/f_{J/\psi}$ as a function of $\Delta t$ are shown in Fig.~\ref{fig:F_th_limit}.
It shows that $\Gamma_{\gamma\nu\bar{\nu}}/f_{J/\psi}$ has an obvious $\Delta t$ dependence, indicating nonnegligible excited-state effects associated with $\phi_{J/\psi}^\dagger$ operator as we have pointed out before.
Using a two-state fit described by Eq.~(\ref{eq:th_fit}) we can extract the ground-state contribution to the ratio at $\Delta t\to\infty$. The results are listed in Table~\ref{tab:width_ratio}.

\begin{figure}[!h]
\centering
\subfigure{\includegraphics[width=0.6\textwidth]{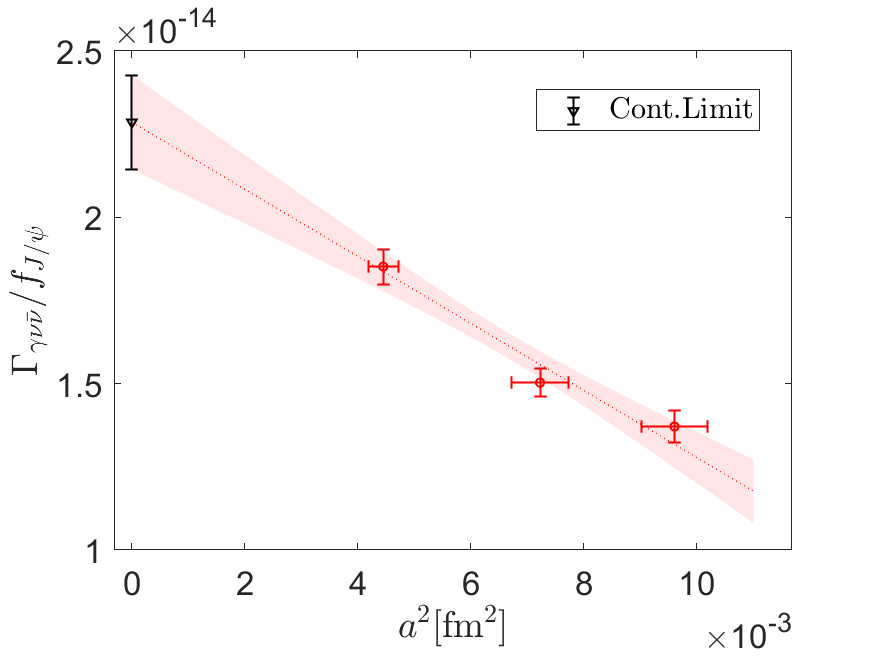}}\hspace{5mm}
\caption{Lattice values of $\Gamma_{\gamma\nu\bar{\nu}}/f_{J/\psi}$ as a function of lattice spacing together with a continuous extrapolation with a linear behavior $a^2$. The errors of lattice spacing are included in the fitting and presented by the horizontal error bars. The symbol of the red circle denotes the lattice results from ensemble a67,a85, and a98 from left to right. The statistical error of $Z_A$ is included by error propagation.}
\label{fig:width_cont_limit}
\end{figure}

\begin{table}[!h]
\centering
\begin{tabular}{cccc}
\hline
\hline
Ensemble & a67 & a85& a98 \\
\hline
$\Gamma_{\gamma\nu\bar{\nu}}/f_{J/\psi}\times 10^{14}$ & 1.852(44) & 1.503(34) & 1.371(25) \\
\hline
\end{tabular}
\caption{Numerical results of $\Gamma_{\gamma\nu\bar{\nu}}/f_{J/\psi}$ for three ensembles.}
\label{tab:width_ratio}
\end{table}

In Fig.~\ref{fig:width_cont_limit}, the lattice results for $\Gamma_{\eta_c\gamma\gamma}/f_{J/\psi}$ at different lattice spacings are shown together with an extrapolation that is linear in $a^2$. We expect this linear behavior since the twisted mass configuration has the so-called automatic $O(a)$ improvement. It is also seen that the fitting curves describe the lattice data well. After the continuous extrapolation, we obtain
$R_f^{\textrm{Cont.Limit}}=2.29(14)\times 10^{-14}$. For a convenient comparison with the experimental branching fraction in the future, we rescale $R_f^{\textrm{Cont.Limit}}$ to physical branching fraction by multipling the $J/\psi$ decay constant $f_{J/\psi}^{\textrm{Cont.Limit}}$ and dividing the total decay width $\Gamma_{J/\psi}=92.6$ keV. Then, the branching factiong is given by $\operatorname{Br}[J/\psi \rw \gamma\nu\bar{\nu}]=1.00(9)\times 10^{-10}$.

Nevertheless, the Ref.~\cite{ETM:2009ptp} claims the ensemble a98 might not be optimally tuned and possibly contain some $\mathcal{O}(a)$ discretization errors. To examine this effect, we also perform our continuum
extrapolation without the coarsest lattice, a98. And then, we get the result $1.07(13)\times 10^{-10}$,
which is consistent with the value $1.00(9)\times 10^{-10}$, but with a larger error.
The consistency suggests there is no residual $\mathcal{O}(a)$ effect on ensemble $a98$.
This conclusion has also been demonstrated in our recent works on charmonium radiative decay~\cite{Meng:2021ecs,Meng:2021las} and
other lattice studies~\cite{ETM:2009ptp,Alexandrou:2009qu,ETM:2009ztk}.
In this paper, we will quote the result with a98 included as the final report and take the difference between these two central values
as our estimation of the systematic error. Our final prediction for the branching fraction of $J/\psi \rw \gamma\nu\bar{\nu}$ is
\be
\operatorname{Br}[J/\psi \rw \gamma\nu\bar{\nu}]=1.00(9)(7)\times 10^{-10}
\ee
where the first error is a statistical error obtained with the spacing error included in the extrapolation and the second is an estimate for the systematic error.

We remark that the relevant phenomenological study in the standard model gives a prediction $\operatorname{Br}[J/\psi \rw \gamma\nu\bar{\nu}]=0.7\times 10^{-10}$~\cite{Gao:2014yga}, which is in the same order of magnitude with our result. Our calculation is performed using three different lattice spacings for the continuous extrapolation, thus the lattice discretization effect is well-controlled. We have also used multiple $\Delta t$ to control the excited-state effects by a multi-state fit. For the neglected disconnected diagrams, they are believed to only give a small contribution
in the charmonium system~\cite{McNeile:2004wu,deForcrand:2004ia,Levkova:2010ft,Hatton:2020qhk} due to the Okubo-Zweig-Iizuka (OZI) suppression.

\section{Conclusion}\label{sec:conclusion}
In this paper, we present a lattice QCD calculation on the invisible decay $J/\psi \rw \gamma\nu\bar{\nu}$ for the first time. 
Our calculation is accomplished using three $N_f=2$ twisted mass fermion ensembles. The excited-state effects are observed and eliminated using a multi-state fit. After a controlled continuous extrapolation, we obtain the first lattice QCD prediction for the branching fraction of $J/\psi \rw \gamma\nu\bar{\nu}$ as $\operatorname{Br}[J/\psi \rw \gamma\nu\bar{\nu}]=1.00(9)(7)\times 10^{-10}$, where the first error is the statistical error that already takes into account the $a^2$-error in the continuous extrapolation, and the second is an estimate of the systematics. The method can also be applied for other processes which involve the leptonic or radiative particles in the final states, for example, $\pi_0 \rw 2\gamma$~\cite{XFeng:2012}, $J/\psi \rw 3\gamma$~\cite{YM:2020} and $K_L\to\mu^+\mu^-$~\cite{NH:lat128}.

Our first-principle calculation provides a precise prediction for the decay of $J/\psi \rw \gamma\nu\bar{\nu}$. It also confirms the previous phenomenological conclusion that the branching fraction of $J/\psi \rw \gamma\nu\bar{\nu}$ is about $10^{-10}$~\cite{Gao:2014yga}. If the future experiments can achieve a precision of $10^{-10}$, the search for new physics scenarios beyond the standard model by the channel $J/\psi \rw \gamma+\textrm{invisible}$ needs to consider the exact contribution of $J/\psi \rw \gamma\nu\bar{\nu}$ from the standard model background.

\begin{acknowledgments}
We thank ETM Collaboration for sharing the gauge configurations with us. We gratefully acknowledge the helpful discussions with Dao-Neng Gao.
Y.M. acknowledges support by NSFC of China under Grant No.12047505 and No.12305094.
The main calculation was carried out on the Tianhe-1A supercomputer at Tianjin National
Supercomputing Center and partly supported by SongShan supercomputer at the National
Supercomputing Center in Zhengzhou.
\end{acknowledgments}

\end{document}